\documentclass[conference]{IEEEtran}
\IEEEoverridecommandlockouts
\usepackage{cite}
\usepackage{amsmath,amssymb,amsfonts}
\usepackage{algorithm}
\usepackage{algpseudocode}
\usepackage{graphicx}
\usepackage{textcomp}
\usepackage{xcolor}
\usepackage{booktabs}
\usepackage{tikz}
\usetikzlibrary{arrows.meta,positioning,fit,backgrounds}
\def\BibTeX{{\rm B\kern-.05em{\sc i\kern-.025em b}\kern-.08em
    T\kern-.1667em\lower.7ex\hbox{E}\kern-.125emX}}

\begin{document}

\title{ECAS: An Edge-Controlled Agentic System for Validation-Gated Scientific Application Execution}

\author{
\IEEEauthorblockN{
Sun Baixi\IEEEauthorrefmark{1},
Mingze Xia \IEEEauthorrefmark{2},
Huihuo Zheng\IEEEauthorrefmark{1}
}
\IEEEauthorblockA{
\IEEEauthorrefmark{1}Argonne National Laboratory, Lemont, IL, USA\\
\IEEEauthorrefmark{2}Oregon State University, Corvallis, OR, USA\\
}
}

\maketitle

\begin{abstract}
Scientific applications increasingly rely on high-performance computing (HPC) systems, yet translating a scientist's high-level goal into a correct target-scale execution remains brittle and labor-intensive. Large language model (LLM) agents offer a promising path toward automating this process, but two obstacles remain. Granting a cloud-hosted model direct HPC system access exposes credentials and execution authority. However, it requires continuous human supervision, limiting automation. Moreover, one-shot generation cannot adapt when generated artifacts fail in a site-specific HPC environment. To address these issues, we present \textsc{ECAS}, an \textbf{E}dge-\textbf{C}ontrolled \textbf{A}gentic \textbf{S}ystem for closed-loop execution of scientific computing campaigns with limited human intervention. \textsc{ECAS} separates \emph{reasoning}, \emph{control}, and \emph{execution}: a cloud-hosted LLM proposes plans, artifacts, and repairs; a user-controlled edge agent retains credentials, workflow state, and execution authority while enforcing policy and resource constraints; and the HPC system performs the computation. Its core mechanism is \emph{validation-gated execution}: device-side generated artifacts for execution on HPC go through static checks and small-scale validation, failures trigger repairs using sanitized execution feedback, and target-scale execution is permitted only after validation and policy checks pass. Additionally, \textsc{ECAS} uses an edge-resident library of expert-distilled, site-specific skills that is never disclosed to the cloud service. In preliminary experiments using three scientific applications on two production ALCF systems with six types of injected faults, closed-loop repair improves application success from 0/6 to 6/6 over one-shot generation. Validation gating prevents all three observed target-scale failures, and skill conditioning improves success from 4/6 to 6/6. These results demonstrate the feasibility of delegating adaptive reasoning to the cloud, while retaining execution control at the edge.
\end{abstract}


\section{Introduction}
Deploying a computational campaign on an HPC system is a translation problem.
A user begins with a goal. For example, ``benchmark the I/O of my data pipeline at scale,'', ``train this model on four nodes'', etc. This goal must be converted into concrete,
system-specific artifacts, including the source or configuration, the correct software
environment (compilers, MPI, Python libraries), a valid launch command for the site's
job launcher, and a batch script with the right queue, account, and resource
request. Small mismatches---an incompatible MPI/Python library, the wrong launcher,
a data path that does not exist on the compute nodes---surface only at runtime,
often after a long queue wait and after consuming allocation.

LLMs are attractive for this translation, and recent systems show they can
generate and even submit HPC and simulation jobs end-to-end~\cite{foamagent,
foamagent2, hpcgpt, hyce}. However, there are two challenges. First, \emph{trust and
reliability}. Concerns arise when pursuing full automation in performing test execution and scale-up: a cloud-hosted model given direct SSH access to a system
would hold user credentials and act on the system with no local checkpoint for
policy, resource, or safety checks, and system logs streamed to the cloud may
leak paths, usernames, and project identifiers. Second, an \emph{adaptation}
obstacle: one-shot generation cannot react to the failures that dominate
real HPC use, where the first artifact is rarely correct, and the informative
signal is the errors in logs from an actual run.

We present \textsc{ECAS} (Edge-Controlled Agentic System), an
agent-based system for automated scientific-computing campaigns that turns a
natural-language goal into a completed, target-scale HPC run as \emph{closed-loop
automation with minimal human intervention}. It is organized around a single
principle: \emph{reason in the cloud, control at the edge, and execute on HPC}.
A cloud-hosted LLM performs open-ended
reasoning---planning, artifact generation, failure diagnosis, and repair
proposals---but never holds credentials or acts directly on the system. A
user-controlled \emph{edge} agent is the trusted control plane: it retains
credentials, workflow state, and artifacts; enforces policy and resource checks;
mediates all system access; sanitizes logs before they leave the trust boundary;
and holds execution authority. The HPC system executes.

On top of this separation, \textsc{ECAS} makes execution \emph{validation-gated}.
Rather than submitting a generated job directly at the requested scale, the edge
agent runs a pipeline: (i) static checks on the artifacts, (ii) a cheap
small-scale or debug-queue validation run, (iii) feedback-driven repair from
sanitized error output, and (iv) a policy-gated promotion to the target-scale run
only after correctness and resource-policy checks pass. This converts expensive
scale-out failures into cheap small-scale failures caught before they consume
allocation.

Closed-loop repair alone, however, is wasteful in a different way: a cold agent
rediscovers the same site-specific fixes on every task---the library-compatible
MPI/Python pairing, the site's launcher and queue conventions---because this
knowledge lives in an expert's head, not in the model. \textsc{ECAS} therefore
treats human expertise as a \emph{build-time} resource: in an offline,
expert-in-the-loop phase an HPC expert works through representative tasks with the
agent, and each correction is distilled into a reusable, edge-private
\emph{skill}---a structured unit of site-specific procedural knowledge that, like
credentials, never leaves the edge. During runtime, the loop is closed and
automated---it retrieves relevant skills to condition generation and repair---and
human effort falls toward zero as coverage grows. Human-in-the-loop is thus our
\emph{building methodology}, not our operating mode: the deployed system runs as
closed-loop automation with minimal human intervention, reduced to an occasional
validator for genuinely novel failures.

This work has three main contributions:
\begin{itemize}
  \item \textbf{A user-controlled edge trust boundary} for LLM-driven HPC
  automation that keeps credentials, state, execution authority, and the
  site-specific skill library off the cloud while still using cloud reasoning,
  and sanitizes feedback crossing the boundary. The cloud proposes only
  abstract, site-agnostic intents; the edge resolves each into a concrete,
  site-correct command through its private skills, so site identifiers and
  execution authority never leave the user's control.
  \item \textbf{A validation-gated execution workflow} that couples static
  checks, small-scale validation, and feedback-driven repair with a policy gate
  before target-scale runs, reducing wasted node-hours.
  \item \textbf{A human HPC expert-in-the-loop skill-building methodology} in which an
  HPC expert works through representative tasks with the agent and each
  expert--agent interaction is distilled into a reusable, edge-private
  \emph{skill}. Using it, we build a concrete site-specific skill library that
  turns an otherwise cold agent into an automated one at deployment and reduces
  repeated repair and expert intervention.
\end{itemize}
We prototype and evaluate \textsc{ECAS} on two production ALCF systems (Crux and Sophia), with
MPI-IO and deep-learning I/O workloads under deterministically injected faults,
quantifying the benefit of closed-loop repair, validation gating, and skill
conditioning.

\section{Background and Related Work}
\textbf{LLMs for HPC.} HPC-GPT adapts an LLM to HPC domain tasks such as
resource and data-race question answering, but targets domain knowledge rather
than a closed execution loop~\cite{hpcgpt}. ``LLM as HPC Expert'' (HyCE) extends
retrieval-augmented generation with hypothetical command embeddings to answer
system-specific questions and retrieve predefined commands, and discusses
data-privacy and command-execution risks, but is retrieval- rather than
execution-centric~\cite{hyce}.

\textbf{Agentic execution and repair.} Foam-Agent and Foam-Agent~2.0 provide an
end-to-end multi-agent pipeline that generates OpenFOAM cases, creates HPC
submission scripts, and monitors and repairs runs, exposing tools via the Model
Context Protocol~\cite{foamagent, foamagent2, mcp}. These systems are the closest
prior art; \textsc{ECAS} differs by treating a user-controlled edge device as the
trust and control plane and by making scale-up explicitly validation-gated for
resource-waste reduction, rather than by targeting one application domain.
Our closed loop follows the reason--act paradigm of ReAct~\cite{react}.

\textbf{Agentic scientific workflows.} Recent work sketches the broader shift to
agentic AI in scientific workflows~\cite{revolution} and proposes reference
architectures and evaluation methodology for LLM agents over workflow
provenance~\cite{provenance}. These are largely conceptual or provenance-focused;
we contribute a concrete edge-controlled prototype and an execution-time
evaluation.

\textbf{Skill libraries and experiential agents.} A line of work lets agents
improve by accumulating reusable experience: Voyager maintains a growing skill
library of executable programs~\cite{voyager}, ExpeL extracts natural-language
insights from past trials~\cite{expel}, and Reflexion converts verbal feedback
into self-improvement signals~\cite{reflexion}. \textsc{ECAS} adapts this idea to
HPC with three differences: skills are sourced from an \emph{HPC expert} in a
build-time loop rather than self-generated; they encode \emph{site-specific}
operational knowledge (modules, launchers, queue/account and ABI conventions);
and they are kept \emph{private on the edge}, since they embed exactly the
identifiers that must not reach the cloud. The library is thus both a performance
mechanism and part of the trust boundary.

\textbf{Workloads and I/O characterization.} We build on standard HPC benchmarks:
IOR for MPI-IO bandwidth~\cite{ior}, the DLIO benchmark for deep-learning I/O
patterns~\cite{dlio}, and E3SM-IO for climate-simulation I/O through the ADIOS2 BP
backend~\cite{e3smio,adios2}, with Darshan for I/O characterization of the
resulting runs~\cite{darshan}.

\begin{figure*}[h]
    \centering
    \includegraphics[width=\linewidth]{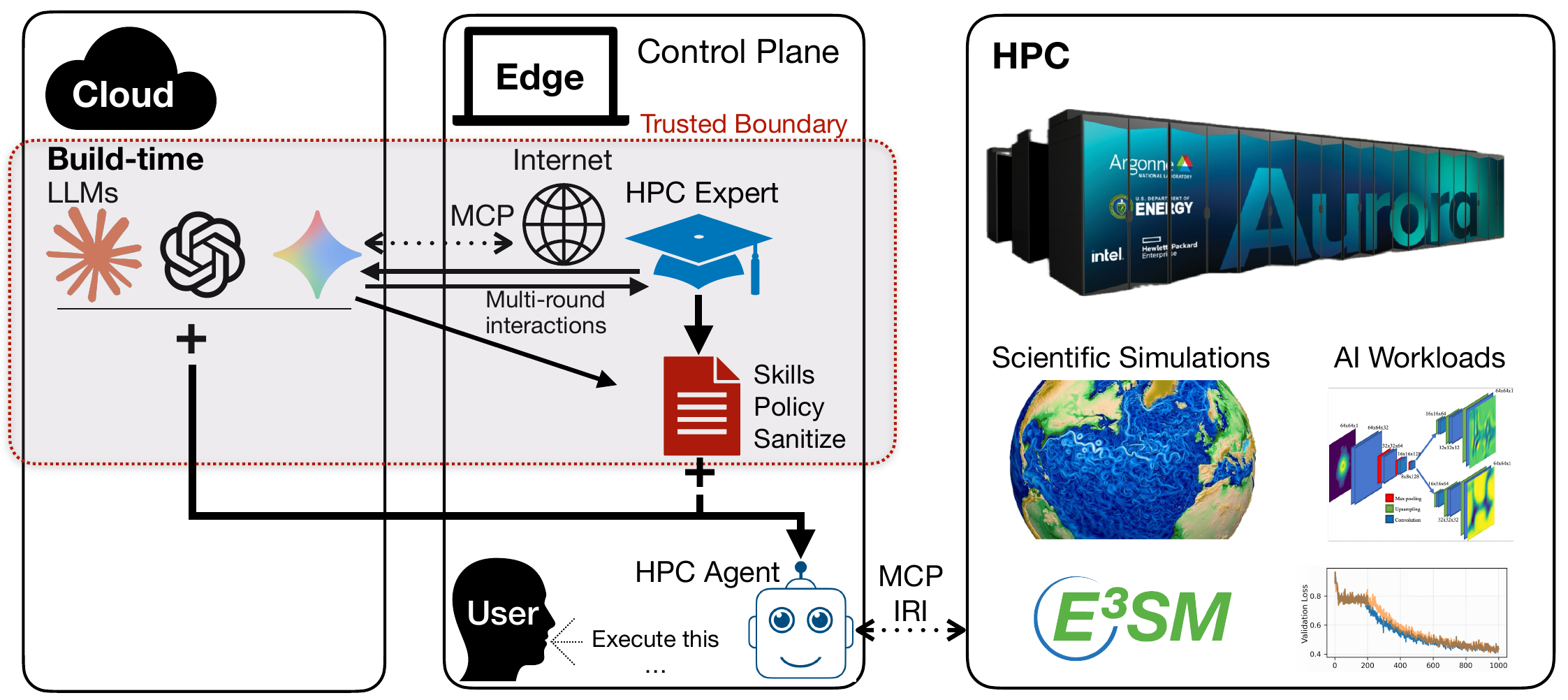}
    \caption{Overview of the ECAS. The edge plays an important role as the Control plane, with the ECAS-built trusted boundary.}
    \label{fig:arch}
\end{figure*}

\section{System Design}

\subsection{Three-Plane Architecture}
\textsc{ECAS} partitions the system into three planes, i.e., edge, cloud, and HPC, with a built-in trusted boundary. The overview of the ECAS system design is presented in Fig.~\ref{fig:arch}.

\emph{Cloud (reasoning plane).} A hosted LLM receives the user goal and sanitized
context and returns structured proposals: an execution plan, artifacts (source or
config, environment specification, launch command, batch script), and, on
failure, an \emph{abstract} diagnosis---a repair category drawn from a fixed,
site-agnostic vocabulary---rather than a site-specific command. Site-specific
values (launcher binaries, absolute paths, account names) are never chosen by the
cloud; the edge resolves an abstract category into a concrete, site-correct fix
using its private skill library (Sec.~\ref{sec:skills}). The cloud plane is
stateless with respect to the system: it holds no credentials and issues no
commands.

\emph{Edge (control plane).} A user-controlled device---the trusted core---owns
everything security- and state-sensitive: user and system credentials; workflow
state and artifact versions; the private skill library (Sec.~\ref{sec:skills});
policy and resource checks; the sole SSH/scheduler connection to the system; log
sanitization; and execution authority. Every action the cloud proposes is
materialized, checked, and (if permitted) executed by the edge agent. Nothing
crosses to the cloud without passing through sanitization.

\emph{HPC (execution plane).} One or more systems execute validation and
target-scale jobs through their batch scheduler via Model Context Protocol (MCP) or Integrated Research Infrastructure (IRI)\cite{DOEIRIPr51:online}. The execution plane trusts only
the edge agent, which authenticates as the user.

\subsection{Threat and Failure Model}
\label{sec:threat}
We assume the cloud LLM is competent but \emph{fallible, not adversarial}: it may
hallucinate or err, i.e., propose the wrong launcher, a nonexistent output path, an
oversized or mis-charged resource request, or an artifact that only fails at
scale. But it does not actively try to exfiltrate data or subvert the edge. The
edge device and its software are trusted and user-controlled, and the HPC site is
trusted to execute jobs submitted as the user. Under this model, the edge's role is
to bound the blast radius of an erroneous cloud proposal with deterministic,
edge-resident checks: (i) static well-formedness of launcher, queue, and account;
(ii) output-path existence and rank/file consistency; (iii) a resource-policy gate
on node count, walltime, queue, account, and remaining node-hour budget
(Eq.~\eqref{eq:policy}); and (iv) sanitization of every log before it crosses the
boundary, stripping usernames, absolute paths, and project identifiers. Because
credentials, workflow state, and the skill library never leave the edge, a cloud
error cannot escalate privilege, spend beyond the user-set policy, or leak site
identifiers; its worst case is a rejected or repaired artifact. We do not defend
against a compromised edge, a malicious site, or side channels within the
sanitized text itself; those are out of scope.

\subsection{Validation-Gated Execution}
The control loop (Algorithm~\ref{alg:vge}) never submits a generated job directly
at target scale. Artifacts first pass \emph{static checks} (e.g., launcher and
queue/account well-formedness, path existence, rank/file consistency). They then
run a \emph{validation} job at minimal scale (one node or a debug queue). Failures
produce sanitized feedback that the cloud turns into a repair; repair is bounded
by fixed budgets on repair attempts, submissions, and wall-clock time. Only after
validation succeeds \emph{and} a resource-policy gate approves the request (nodes,
walltime, account) is the job promoted to the target-scale run. Success is judged
by an application-specific oracle, not by the scheduler's \texttt{COMPLETED}
status alone.

We make the edge's control decision explicit. A candidate artifact set $a$ carries
a resource footprint $\rho(a)=(n(a),w(a),q(a),\alpha(a))$: node count, wall-clock
request, queue, and charging account. The edge holds a site policy
$\Pi=(N_{\max},W_{\max},\mathcal{Q},\mathcal{A},H_{\mathrm{rem}})$---per-job node and
walltime ceilings, admissible queues and accounts, and the remaining node-hour
budget---and the gate is the conjunction of deterministic checks
\begin{equation}
\label{eq:policy}
\mathrm{PolicyGate}(a)=\prod_{\phi\in\Phi}\mathbf{1}\!\left[\phi(a)\right],
\end{equation}
with $\Phi=\{\,n\!\le\!N_{\max},\; w\!\le\!W_{\max},\; q\!\in\!\mathcal{Q},\;
\alpha\!\in\!\mathcal{A},\; n\,w\!\le\!H_{\mathrm{rem}}\,\}$, evaluated entirely on the
edge. Target-scale execution is authorized only when the validation run passes the
oracle \emph{and} the target footprint clears the gate,
\begin{equation}
\label{eq:promote}
\mathrm{Promote}(a)\;\Longleftrightarrow\;
\mathrm{Oracle}(r_{\mathrm{val}})\;\wedge\;\mathrm{PolicyGate}(a_{\mathrm{target}}).
\end{equation}
Equivalently, target-scale execution is allowed iff three conditions hold
independently at the edge: the validation run passed its oracle
($\mathrm{ValidationPassed}$); the request is policy-admissible
($n\!\le\!N_{\max}$, $w\!\le\!W_{\max}$, $q\!\in\!\mathcal{Q}$,
$\alpha\!\in\!\mathcal{A}$; $\mathrm{PolicySatisfied}$); and it fits the residual
allocation ($n\,w\!\le\!H_{\mathrm{rem}}$; $\mathrm{ResourceBudgetSatisfied}$).
Because $\Pi$ and both predicates reside at the edge, neither a cloud plan nor a
generated artifact can escalate scale or spend beyond the user-set policy.

\begin{algorithm}[t]
\caption{Validation-gated execution at deployment (edge agent). The skill library
$S$ is built offline (Sec.~\ref{sec:skills}).}
\label{alg:vge}
\begin{algorithmic}[1]
\Require goal $g$; skill library $S$; budgets $B_{\text{rep}}, B_{\text{sub}}, B_{\text{time}}$
\State $k \gets \mathrm{Retrieve}(S, g)$ \Comment{relevant skills}
\State $a \gets \mathrm{CloudPlanAndGenerate}(g, k)$ \Comment{skill-conditioned artifacts}
\For{$i = 1$ \textbf{to} $B_{\text{rep}}$}
  \If{$\lnot\ \mathrm{StaticCheck}(a)$}
    \State $a \gets \mathrm{CloudRepair}(a, \text{static diag.}, k)$
    \State \textbf{continue}
  \EndIf
  \State $r \gets \mathrm{RunValidation}(a)$ \Comment{1 node / debug}
  \If{$\mathrm{Oracle}(r)$ \textbf{and} $\mathrm{PolicyGate}(a_{\text{target}})$}
    \State \textbf{return} $\mathrm{RunTargetScale}(a)$
  \EndIf
  \State $a \gets \mathrm{CloudRepair}(a, \mathrm{Sanitize}(r), k)$
\EndFor
\State \textbf{return} \textsc{Failed} \Comment{escalate to expert}
\end{algorithmic}
\end{algorithm}

\subsection{Skill-Building Methodology}
\label{sec:skills}
\textsc{ECAS} operates in two phases. The deployment loop above is automated, but
it is only effective once the edge holds site-specific knowledge that a
general-purpose cloud model lacks. We obtain that knowledge \emph{offline}, with
a human in the loop, and then reuse it automatically.

\emph{Build phase (expert-in-the-loop).} An HPC expert works through a set of
representative tasks with the agent. When the automated loop stalls---budget
exhausted, or a repair that keeps failing the same way---or when it produces an
artifact that runs but violates a site convention, the expert supplies a
correction. The edge distills the interaction into a \emph{skill}: a structured,
retrievable record pairing a trigger (workload, site, and error signature) with
the procedural knowledge that resolved it---an ABI-compatible module set, a
launcher template, a queue/account rule, or a known-good environment recipe.
Skills are versioned alongside artifacts and stored in a library on the edge.
Because a skill embeds exactly the usernames, paths, and project conventions that
sanitization strips, the library is inherently edge-private; it is never shipped
to the cloud, and only the retrieved skill's sanitized guidance conditions a
prompt.

\emph{Deployment phase (automated).} For a new goal, the edge retrieves relevant
skills and prepends their guidance to cloud generation and repair
(Algorithm~\ref{alg:vge}). Tasks covered by the library are handled end to end
with no human involvement; a genuinely novel task that exhausts its budget
escalates back to the expert, whose correction seeds a new skill. The system thus
degrades gracefully and improves monotonically: human effort is spent once per
class of failure, not once per run.

\section{Implementation}
The edge agent is a Python controller that holds credentials, drives the system
over SSH and the PBS scheduler, versions artifacts, and applies static, policy,
and sanitization steps. The reasoning plane is a hosted LLM accessed through an
institutional inference gateway. We deploy on two production ALCF systems:
\emph{Crux} (PBS Pro, Cray MPICH) and \emph{Sophia} (NVIDIA DGX-A100, OpenMPI).
Three workloads exercise the pipeline: IOR for MPI-IO bandwidth~\cite{ior}, the
DLIO benchmark for deep-learning I/O~\cite{dlio}, and E3SM-IO~\cite{e3smio} for
climate-simulation I/O written through the ADIOS2 BP backend~\cite{adios2};
Darshan captures I/O behavior of each run~\cite{darshan}. E3SM-IO doubles as a
\emph{cross-system portability} case: the same source is built and run on both
Crux (Cray MPICH) and Sophia (OpenMPI), whose incompatible MPI ABIs mandate
independent builds and site-specific launchers.
Application-specific oracles define success (e.g., IOR reports nonzero aggregate
bandwidth over all segments and a Darshan log is produced; DLIO completes the
configured steps and reports throughput; E3SM-IO exits cleanly and reports
nonzero write bandwidth for its ADIOS2 BP output). Static checks and the policy gate are
deterministic and run entirely on the edge; log sanitization strips usernames,
absolute paths, and project identifiers before any text is sent to the cloud.
The skill library is a set of structured records on the edge, keyed by workload
and site and retrieved by matching the goal and any error signature; retrieval
prepends the matched skills' guidance to the cloud prompt, so no library content
leaves the trust boundary in raw form.

\section{Evaluation}
\begin{table}[t]
\caption{Experimental configuration.}
\label{tab:config}
\centering
\footnotesize
\begin{tabular}{ll}
\toprule
Parameter & Value \\
\midrule
Systems & Crux (PBS Pro, Cray MPICH, PALS \texttt{mpiexec}, \\
        & 256 cores/node); Sophia (OpenMPI, 8$\times$A100) \\
Workloads & IOR (MPI-IO), DLIO, E3SM-IO (ADIOS2 BP) \\
Validation scale & 1 node (IOR/DLIO); 16-proc, 1 node (E3SM-IO) \\
Target scale & 4 nodes (IOR/DLIO); 512-proc, 2 nodes (E3SM-IO) \\
Injected faults & 6 (2 static-decidable, 4 runtime; 2 site-specific) \\
Repair budget & $B_{\text{rep}}=3$ attempts \\
Reasoning model & hosted LLM via institutional inference gateway \\
Repetitions & $N=5$ per fault per condition (E2, E4) \\
\bottomrule
\end{tabular}
\end{table}
\subsection{Experimental Setup}
We evaluate ECAS on two systems using three workflows, as presented in Table \ref{tab:config}.

\emph{Systems.} We evaluate ECAS on two production systems at the Argonne Leadership Computing
Facility (ALCF). Crux is a CPU-based system running the PBS Pro scheduler with the
Cray MPICH stack and the PALS mpiexec launcher, providing 256 cores per node; it
hosts the end-to-end validation and target-scale runs. Sophia is a GPU system
with OpenMPI and 8 NVIDIA A100 GPUs per node, used to exercise a distinct site
environment (different launcher and MPI conventions).

\emph{Workloads and scales.} Three standard scientific-computing workloads span the I/O regimes of interest:
IOR for MPI-IO bandwidth, the DLIO benchmark for deep-learning I/O patterns, and
E3SM-IO for climate-simulation I/O through the ADIOS2 BP backend. Each workload
is exercised under the validation-gated pipeline at two scales: a cheap
validation scale (1 node for IOR/DLIO; 16 processes on 1 node for E3SM-IO) at
which failures are caught before they consume allocation, and a target scale
(4 nodes for IOR/DLIO; 512 processes on 2 nodes for E3SM-IO) to which a run is
promoted only after it passes validation and the resource-policy gate.

\emph{Injected faults and repair loop.} To probe the closed loop, we inject six fault types that mirror the mismatches
that dominate real HPC use: two are statically decidable (caught by artifact
checks before submission) and four surface only at runtime; two of the six are
site-specific (resolvable only through the edge's private skill library, not by
generic cloud reasoning). The repair loop is bounded by a budget of B\_rep = 3
attempts per task before escalation to a human.

\emph{Reasoning model and repetitions.} Failure diagnosis is produced by Claude-opus-5 \cite{Introduc2:online} accessed through the institution's
inference gateway; the edge maps the model's abstract diagnosis to a concrete,
site-correct fix. To capture the variance of a real LLM in the loop, every fault
is run N = 5 times per condition in experiments E2 and E4.

\emph{Research Questions and Evaluation Structure} We ask four questions. \textbf{(RQ1)} Can \textsc{ECAS} take a natural-language
goal to a completed target-scale run on real workloads? \textbf{(RQ2)} How much
does closed-loop repair improve over one-shot generation under realistic faults?
\textbf{(RQ3)} How much target-scale waste does validation gating avoid?
\textbf{(RQ4)} Does build-time skill conditioning reduce repeated repairs and
expert intervention at deployment? 
All experiments use fixed repair, submission, and time budgets and
application-specific oracles. To control for scheduler-queue variability, we report
\emph{queue-excluded time} (wall-clock minus PBS queue wait, from scheduler
timestamps) alongside counts.

\emph{Metrics.} For a fault set $F$, let $o_f\!\in\!\{0,1\}$ be the oracle outcome
for fault $f$, $e_f$ the number of expert interventions it requires, and $R_f,S_f$
its repair and submission counts. A run counts as a success only when all four
checks pass---the scheduler reports the job completed
($\mathrm{SchedPass}$), the application exits zero ($\mathrm{ExitPass}$), the
workload's application-level metric is present and nonzero ($\mathrm{AppOracle}$;
IOR aggregate bandwidth, DLIO throughput, or E3SM-IO write bandwidth), and the
expected output artifact exists and is well-formed ($\mathrm{ArtifactOracle}$;
a Darshan log, or \texttt{bpls} listing the expected variables in the ADIOS2 BP
output):
\begin{equation}
\label{eq:oracle}
o_f=\mathrm{SchedPass}\wedge\mathrm{ExitPass}\wedge\mathrm{AppOracle}\wedge
\mathrm{ArtifactOracle}.
\end{equation}
This composite oracle is what distinguishes a genuinely completed run from one the
scheduler merely marks \texttt{COMPLETED}. We report success rate $\mathrm{SR}$, autonomous
completion $\mathrm{AC}$ (resolved with no expert), and mean repairs and
submissions,
\begin{equation}
\label{eq:rates}
\begin{aligned}
\mathrm{SR}&=\tfrac{1}{|F|}\!\sum_{f\in F}\! o_f, &
\mathrm{AC}&=\tfrac{1}{|F|}\!\sum_{f\in F}\!\mathbf{1}[o_f{=}1\wedge e_f{=}0],\\[2pt]
\bar{R}&=\tfrac{1}{|F|}\!\sum_{f} R_f, &
\bar{S}&=\tfrac{1}{|F|}\!\sum_{f} S_f.
\end{aligned}
\end{equation}
Node-hours are attributed by stage. With validation at $n_{\mathrm v}$ nodes and
target scale $n_{\mathrm t}$, the wasted (non-successful) node-hours of a strategy are
\begin{equation}
\label{eq:waste}
H_{\mathrm{waste}}=\sum_{f\in F} n_{\mathrm v}\,t^{\mathrm v}_f
\;+\!\!\sum_{f:\,o_f=0}\!\! n_{\mathrm{fail}(f)}\,t^{\mathrm{fail}}_f ,
\end{equation}
where $n_{\mathrm{fail}(f)}$ is the scale at which fault $f$ fails:
$n_{\mathrm{fail}}{=}n_{\mathrm t}$ under direct submission (which performs no
validation, so its first term vanishes) and $n_{\mathrm{fail}}{=}n_{\mathrm v}$ under
validation gating. Gating thus scales each failure's wasted node-hours by
$n_{\mathrm v}/n_{\mathrm t}$, so its advantage grows with target scale and
fault-detection latency.

\subsection{E1: End-to-End Feasibility (RQ1)}
We run the full pipeline for all three workloads (IOR, DLIO, and E3SM-IO),
validating at small scale on one node and promoting to a target run. IOR and
DLIO promote to a four-node target; E3SM-IO, whose F-case decomposition map pins
the process count, validates with the 16-process map on one node and promotes to
the 512-process map on two nodes. We report end-to-end completion, agent turns,
and oracle pass/fail (Table~\ref{tab:e1}). All three complete on Crux: E3SM-IO
writes its F-case history files through the ADIOS2 BP backend at both scales,
reaching $1112$~MiB/s aggregate write bandwidth at the 512-process target, and
the output passes the artifact oracle---a \texttt{bpls} listing confirms all
$162$ F-case variables in the resulting BP dataset. To exercise the cross-system
portability path, E3SM-IO and its ADIOS2/PnetCDF stack are additionally built
and link-verified on Sophia under OpenMPI, confirming that the same deployment
pipeline compiles and links against an incompatible MPI ABI (Cray MPICH vs.\
OpenMPI); we report the Crux end-to-end runs in Table~\ref{tab:e1}.

\begin{table}[t]
\caption{E1: end-to-end feasibility on Crux. One-node validation is
promoted to the target scale (four nodes for IOR/DLIO; the map-pinned
512-process, two-node configuration for E3SM-IO). E3SM-IO is additionally
built and link-verified on Sophia under OpenMPI (cross-system portability).}
\label{tab:e1}
\centering
\begin{tabular}{llccc}
\toprule
Workload & System & Completed & Agent turns & Oracle \\
\midrule
IOR (MPI-IO)     & Crux   & Yes & 2 & pass \\
DLIO             & Crux   & Yes & 2 & pass \\
E3SM-IO (ADIOS2) & Crux   & Yes & 2 & pass \\
\bottomrule
\end{tabular}
\end{table}

\subsection{E2: One-Shot vs.\ Closed-Loop Agent (RQ2)}
We inject six deterministic faults into the initial state and compare a
\emph{one-shot} baseline (generate once, submit, no repair) with the
\emph{closed-loop} agent. The faults, chosen to span static- and runtime-detected
failures, are: (1) wrong queue/account; (2) transfer size exceeding block size
(a static-decidable IOR misconfiguration); (3) invalid output path; (4) wrong
launcher (OpenMPI \texttt{mpirun} in place of the Cray PALS \texttt{mpiexec});
(5) excessive ranks-per-node that overruns the launcher's local-rank limit; and
(6) an MPI ABI mismatch (an OpenMPI library \texttt{LD\_PRELOAD}ed into Cray
MPICH). Faults (1)--(2) are caught by static checks; (3)--(6) surface only at
runtime, and (4) and (6) additionally require site-specific knowledge to repair.
Because the reasoning model is stochastic, we repeat every fault $N{=}5$ times per
condition and report successes with denominators (Table~\ref{tab:matrix}); the
one-shot baseline is a single attempt per fault. The one-shot baseline resolves
$0/6$ faults; the closed-loop agent (full \textsc{ECAS}, skill column) resolves
every fault in every repetition ($30/30$ runs, all autonomous) at a mean of
$1.00$ repairs and $2.67$ submissions per resolved run (one-shot: $0.67$
submissions, no repair).

\begin{table}[t]
\caption{E2/E4: per-fault oracle outcomes over $N{=}5$ repetitions per
condition. \emph{One-shot}: single generate-and-submit, no repair (E2
baseline). \emph{Cold}: closed-loop repair with an empty skill library (E4).
\emph{Skill}: closed-loop repair with the edge skill library---the full
\textsc{ECAS} system (E2 closed-loop; E4 skill-conditioned). Entries are
successes over five repetitions; $\dagger$ marks faults that require
site-specific knowledge to repair.}
\label{tab:matrix}
\centering
\begin{tabular}{lccc}
\toprule
Injected fault & One-shot & Cold & Skill \\
\midrule
Wrong queue/account        & $\times$ & 5/5 & 5/5 \\
Transfer $>$ block size    & $\times$ & 5/5 & 5/5 \\
Invalid output path        & $\times$ & 5/5 & 5/5 \\
Wrong launcher$^\dagger$   & $\times$ & 0/5 & 5/5 \\
Excess ranks/node          & $\times$ & 5/5 & 5/5 \\
MPI ABI mismatch$^\dagger$ & $\times$ & 0/5 & 5/5 \\
\midrule
Total (successful runs)    & 0/6 & 20/30 & 30/30 \\
\bottomrule
\end{tabular}
\end{table}

\subsection{E3: Direct vs.\ Validation-Gated Scale-Up (RQ3)}
For three faults that pass static checks but fail at runtime (invalid output
path, wrong launcher, and excessive ranks-per-node), we compare \emph{direct}
target-scale submission (four nodes) against \emph{validation-gated} execution
(validate at one node, repair, then four nodes), repeating each fault $N{=}5$
times per strategy (15 real end-to-end runs each) with the same closed loop. We
account node-hours by where they are spent---validation (one node), failed
attempts, and successful runs---and count failed runs by the scale at which they
occur (Table~\ref{tab:e3}). Because these faults fail fast, the non-productive
node-hours are comparable across strategies ($0.030$ vs.\ $0.036$); the decisive
difference is \emph{scale}. Direct submission incurs a four-node target-scale
failure in \emph{every} one of its 15 runs ($15/15$), whereas gating confines
every failure to one-node validation, leaving zero failed target-scale runs
($0/15$). The node-hour advantage of gating therefore grows with fault-detection
latency and target scale, which our short-running, small-scale faults
deliberately understate.

\begin{table}[t]
\caption{E3: node-hour accounting and failure scale over three runtime faults,
summed across $N{=}5$ repetitions (15 runs per strategy).}
\label{tab:e3}
\centering
\begin{tabular}{lcc}
\toprule
 & Direct & Validation-gated \\
\midrule
Validation node-hours (1-node)   & 0        & 0.036 \\
Failed node-hours (4-node)       & 0.030    & 0 \\
Successful node-hours (4-node)   & 0.319    & 0.318 \\
\midrule
Non-productive node-hours        & 0.030    & 0.036 \\
Failed \emph{target-scale} runs  & 15/15    & 0/15 \\
\bottomrule
\end{tabular}
\end{table}

\subsection{E4: Cold vs.\ Skill-Conditioned Agent (RQ4)}
We first build a skill library offline: an HPC expert works through a training set
of tasks and faults, and each unresolved or site-suboptimal case is distilled
into a skill (Sec.~\ref{sec:skills}). We then run the automated deployment loop on
the six E2 faults under two conditions: \emph{cold} (empty library) and
\emph{skill-conditioned}. Two of the six faults (wrong launcher, MPI ABI
mismatch) require site-specific knowledge to repair; a general agent without the
library cannot resolve them within the repair budget and escalates to a human.
Over the six faults and five repetitions (Table~\ref{tab:matrix}, cold vs.\
skill), the cold agent resolves the four non-site-specific faults every time
($20/30$, $67\%$) but never either site-specific fault ($0/10$), escalating to a
human in all ten cases. The skill-conditioned agent resolves all six every time
($30/30$, $100\%$), including $10/10$ on the site-specific faults, with
\emph{zero} expert interventions. The skill library does not reduce repair count
(both average $1.00$) but converts the two otherwise-unresolvable faults into
autonomous fixes.

\section{Discussion and Limitations}
This is a research prototype and the results are preliminary. Our evaluation uses
two related ALCF systems, a single inference gateway, deterministic (not
adversarial) fault injection, and small target scales; the faults are
representative but not exhaustive. We fix repair, submission, and time budgets,
so reported successes are budget-conditioned. We do not evaluate broad security
properties, and we deliberately avoid claiming the system is ``secure,''
``production-scale,'' or ``large-scale.'' Our automation claim is scoped: we
describe \textsc{ECAS} as closed-loop automation with \emph{minimal human
intervention}, not as ``fully autonomous.'' The deployment loop runs automatically
only for task classes the skill library covers, and genuinely novel tasks still
escalate to the expert. Skills are built by a single expert on a small task set and
may encode site- or person-specific idiosyncrasies; retrieval can misfire, and
skill quality bounds the benefit we report. Log sanitization reduces but does not
eliminate leakage risk, and the policy gate encodes only the checks we implement.
Generalization of both the framework and the skill library across schedulers,
sites, experts, and models is future work.

\section{Conclusion}
We presented \textsc{ECAS}, an edge-controlled agentic system for automated
scientific-computing campaigns that separates cloud reasoning, edge control, and
HPC execution, and gates target-scale execution behind static checks, small-scale
validation, and feedback-driven repair. It takes a natural-language goal to a
completed target-scale run as closed-loop automation with minimal human
intervention. By keeping credentials, state, the skill library, and execution
authority on a user-controlled edge and promoting jobs only after validation,
\textsc{ECAS} provides credential isolation and policy-mediated execution for
LLM-driven HPC automation while reducing wasted node-hours. A
build-time, expert-in-the-loop methodology captures site expertise once as
reusable skills, so the deployed loop runs automatically and needs the expert
only for genuinely new failures. Our preliminary prototype on two ALCF systems
suggests closed-loop repair, validation gating, and skill conditioning
meaningfully improve success and reduce wasted node-hours; a full evaluation is
ongoing.

\bibliographystyle{IEEEtran}
\bibliography{references}

\end{document}